# Samuel Preston and *E=mc²*

Jean-Paul Auffray

*07700 Saint-Martin d'Ardèche, France*
(e-mail: jpauffray@yahoo.fr)

**Abstract**. In his 1875 determination of the mass-energy equivalence, did Samuel Preston actually consider ether to be "rarefied mass" and mass to be "concentrated ether"?

**Keywords:** Preston, ether, mass, energy, E=mc2, Poincaré.

## INTRODUCTION

In a recent communication diffused on arXiv [1], I briefly alluded to Samuel Tolver Preston's 1875 reasoning which led him to the relation E=mc² *via* Le Sage's postulate of the existence in space of tenuous ether particles moving at the speed of light. Dr Kyriazi of the University of Pittsburgh School of Medicine has raised the question as to whether Preston's line of reasoning actually relies on the assumption that ether is "rarefied mass" and mass "concentrated ether", as I suggested in my communication [2]. The present communication addresses this question.

## 1. PRESTON'S 1875 POSTULATES

Samuel Preston's starting point is unambiguous. Rejecting the concept of "potential energy", which he considered to be as mythical as that of "action at a distance", Preston asserted that energy exists only in the form of motion – in modern terms, as *kinetic* energy – thus recovering Leibniz's concept of a *vis motrix* proportional to the square of velocity. Assuming purely mechanical processes to explain the workings of nature, Preston then postulated, following Le Sage, the existence in space of ether particles moving at the speed of light and exerting a pressure against the molecules of matter. On this basis, Preston reached the conclusion that "the above deduction, as to the high speed of the ether particles in their normal state, throws at once a light upon the existence of a vast store of energy in space of a very intense character". This is the crux of the line of reasoning which led Preston to formulate a mass-energy relation [3].

## 2. ETHER, AS A STORE OF ENERGY

Preston envisioned the possibility of "interchange of motion between the ether and the molecules of matter under special conditions". Speculating as to what these "special conditions" might be, he imagined a process in which a given "molecule of matter" subdivides into an increasing number of "parts" or "particles", each particle keeping the same amount of (kinetic) energy the original molecule possessed. This can occur only if each particle thus generated acquires an increased velocity to compensate for its diminished mass. Preston noted that, "being subdivided among a large number of particles", the total energy the subdivided molecule originally held becomes ultimately "vastly greater". He concluded: "We might imagine this process to go on progressively until at length the dimensions, mean distance, and speed of the ether particles themselves had been reached."

This statement may be construed to signify that, in Preston's view, a molecule of matter can subdivide into a large number of ether-like particles, suggesting that ether is, indeed, some kind of "rarefied mass".

The relevance of this interpretation is reinforced when one considers how Preston dealt with the following statement: "We shall now consider more particularly the energy enclosed by the ether." He noted: "This requires […] first, a knowledge of *the quantity of matter in the form of ether* [my emphasis] contained in the unit volume of space (i. e. the density of the ether)."

Thus, in Preston's view, matter appears to be some form of "condensed ether" [4].

Whether or not these interpretations of Preston's nineteenth century language are fully justified, remains to be ascertained. It remains a fact, however, that Samuel Preston based his search of an "equivalence" between mass and energy on considerations involving the postulated existence in nature of "ether particles" possessing a large amount of (hidden) quantity of motion, or, equivalently, possessing a large (hidden) quantity of (kinetic) energy.

# 3. PRESTON'S MASS-ENERGY RELATION

Using *ad hoc* arguments, Preston estimated the pressure ether particles supposedly exert on matter molecules to be at least equal to "500 tons per square inch". He established his mass-energy relation on the basis of this estimate, reaching this conclusion: "A quantity of matter representing a total mass of only one grain, and possessing the normal velocity of the ether particles (that of a wave of light), encloses a store of energy represented by upwards of one thousand millions of foot-tons."

Does this statement correspond precisely to the equation $E=mc^2$?

When mass is measured in grains [5], energy in foot-pounds, and velocity in feet per second, the kinetic energy of a mass of one grain travelling at the speed c is equal to $c^2$ multiplied by 450 395.

One foot-ton equals 2240 foot-pounds. A kinetic energy of "one thousand millions of foot-tons" corresponds therefore to $2,24 \times 10^{12}$ foot-pounds.

The speed of light is approximately equal to $9 \times 10^8$ feet per second. Its square is therefore approximately equal to $81 \times 10^{16}$ feet per second squared. Divided by 450 395, this gives a value of approximately $1,8 \times 10^{12}$ foot-pounds for the mass of one grain expressed in energy units. This value may be compared to the value of $2,24 \times 10^{12}$ foot-pounds obtained above for the kinetic energy. Preston's estimate thus corresponds, to a good approximation, to the formula $E-mc^2$, an amazing result.

Preston did not envision the possibility that matter could be "destroyed' to release energy, as modern physicists do. This is clearly indicated in his remark that "the tremendous energy developed in explosives, *which is the very energy of the ether itself* [my emphasis], is a direct indication of the ether pressure, which is the necessary accompaniment of this energy".

Preston well understood the tremendous implications of his ether theory: "A quantity of matter representing a mass of one grain *endued with the velocity of the ether particles* [my emphasis], encloses an amount of energy which, *if entirely utilized* [my emphasis], would be competent to project a weight of one hundred thousands tons to a height of nearly two miles (1.9 miles)."

# 4. CONCLUSION

The suggestion that, in 1875, Samuel Preston (at least implicitly) held the view that ether is "rarefied mass" and mass "concentrated ether" appears to be not altogether incorrect. This suggestion is only marginally relevant, however, to the line of reasoning which led Preston to estimate, with an amazing foresight, the quantity of energy "contained" or "enclosed" in a given space in the form of ether particles moving at high speed. I am grateful to Dr Kyriazi for having called my attention to the interesting questions my casual original remark concerning Preston's contribution to the discovery of $E=mc^2$ raises.

# REFERENCES

1. J.-P. Auffray, *Dual origin of* E-mc$^2$, arXiv doc. Gen-phys 0608289 (2006).
2. Harold Kyriazi, Dept. of Neurobiology, University of Pittsburgh School of Medicine, Pittsburgh, PA 15261, private communication.
3. All quotes presented in this paper are taken from Samuel Tolver Preston's book, *The Physics of Aether* (1875).
4) Preston estimated the density of air to be 5 264 800 times greater than that of ether.
5) One "grain" equals about 64.8 milligrams.